\documentclass[twocolumn,superscriptaddress,prl]{revtex4-2}

\usepackage{amsmath}
\usepackage{amssymb}  
\usepackage{braket}
\usepackage{graphicx}
\usepackage{subcaption}
\usepackage{xcolor}
\usepackage{ulem}  
\usepackage[colorlinks=true,linkcolor=blue,citecolor=blue,urlcolor=blue]{hyperref}  
\usepackage{todonotes}
\usepackage{nicefrac}  
\usepackage{float}

\begin{document}

\begin{abstract}
Contact processes play an important role in classical non-equilibrium dynamics, describing the spreading of diseases, the dynamics of earthquakes and forest fires, and the distribution of information through the internet. 
Here we show that their quantum counterpart, where the spreading occurs through coherent couplings, displays even richer dynamics and offers new means of control. 
A quantum contact process on a topologically non-trivial lattice can be confined to a protected subspace corresponding to 
either a single site or a fully excited lattice.
Furthermore, excitation spreading can be controlled to occur in quantized steps and on demand when employing topological pumps. 
We show that the many-body dynamics of excited domains can be mapped to an effective single-particle model, which also determines the topological properties.
Throughout this work, we consider a specific type of contact process corresponding to 
coherent Rydberg facilitation in a tweezer array of trapped atoms in a one-dimensional lattice.
\end{abstract}

\title{Quantum Contact Processes on a Topological Lattice}

\author{Julius Bohm} 
\affiliation{Department of Physics and Research Center OPTIMAS, RPTU University Kaiserslautern-Landau, D-67663 Kaiserslautern, Germany}
\author{Richard Schmidt}
\affiliation{Institut für Theoretische Physik, Universit\"at Heidelberg, 69120 Heidelberg, Germany}
\author{Michael Fleischhauer}
\affiliation{Department of Physics and Research Center OPTIMAS, RPTU University Kaiserslautern-Landau, D-67663 Kaiserslautern, Germany}
\affiliation{Research Center QC-AI, RPTU University Kaiserslautern-Landau, 67663 Kaiserslautern, Germany}
\author{Daniel Brady}
\affiliation{Department of Physics and Research Center OPTIMAS, RPTU University Kaiserslautern-Landau, D-67663 Kaiserslautern, Germany}

\date{\today}

\maketitle

\paragraph{Introduction -- }

Spreading phenomena such as epidemics or the flow of information are described by classical contact processes. One of their characteristic features is the competition between excitation spreading and dissipation. 
This results in non-equilibrium phase transitions between an absorbing state of vanishingly small activity and an active phase of permanent excitation spreading, generically belonging to the directed percolation (DP) universality class \cite{hinrichsen2008book}. 
Apart from facilitating such a phase transition, there is, however, little control over the spreading process. 

In its quantum counterpart, the quantum contact process (QCP) \cite{carollo2019critical}, the spreading dynamics can be much 
richer due to interference phenomena associated with \textit{coherent} transport. As a consequence, the QCP can display dynamic behavior, which does not exist in classical spreading processes, such as Bloch oscillations \cite{magoni2021emergent}. Moreover, quantum fluctuations become important and may 
modify critical phenomena and the universal properties of the absorbing-state phase transition \cite{marcuzzi2016absorbing,buchhold2017nonequilibrium,gillman2019numerical,carollo2019critical}.

\begin{figure}[H]
  \centering
  \includegraphics[width=0.95\columnwidth]{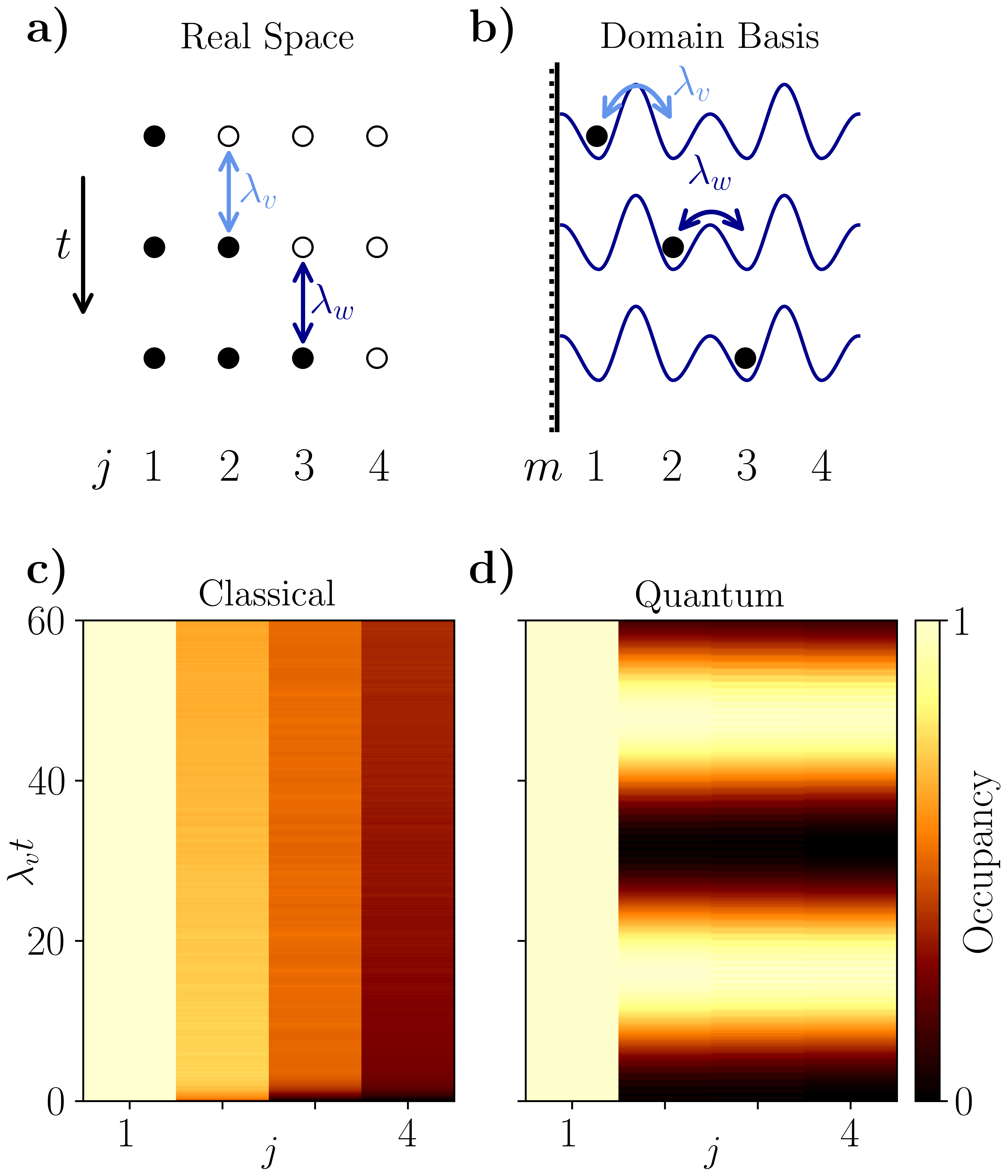}
  \caption{\textit{QXP quantum contact process on a one-dimensional lattice.} (a) In real space $\lambda_v = 1$ and $\lambda_w = 10$ denote alternating coherent excitation rates (Rabi frequencies) on even and odd sites. (b) In the basis of domains of length $m$ domain space these correspond to alternating hopping rates. 
  Time and space resolved dynamics for the classical (c) and quantum (d) QXP contact process for four sites with open boundary conditions and with one initial seed excitation. The color code represents the excitation probability.
  }
\label{fig:intro}
\end{figure}

In this letter, we show that in the quantum case, spreading processes can be controlled by topology. They can be confined to topologically protected subspaces and can proceed in quantized steps controlled by topological invariants. 
For this, we consider a specific model of the QCP in one dimension. 
Here, each site can either be excited $\ket{\bullet}$ or in its ground state $\ket{\circ}$. Specifically, we focus on the QXP model, where excitations can only be created if exactly \textit{one} neighboring site is already excited, i.e. $\ket{\bullet \circ \circ} \leftrightarrow \ket{\bullet \bullet \circ}$, while they are suppressed if either \textit{none} or \textit{two} neighbors are excited, i.e. $\ket{\circ\!\circ\!\circ} \not\leftrightarrow \ket{\circ\!\bullet\!\circ}$ and $\ket{\bullet\!\circ\!\bullet} \not\leftrightarrow \ket{\bullet\!\bullet\!\bullet}$ (cf. Fig.~\ref{fig:intro}).

The kinetic constraint on the many-body dynamics makes the QXP model the quantum counterpart to classical epidemic models, including the concrete 
 implementation of an initial seed (''patient zero'').  
%
%
\noindent 
It is a variant of the PXP model \cite{bak1982one,fendley2004competing,burnell2009devil},  a paradigmatic quantum system
with kinetic constraints, where excitation is only possible in the absence of any excited (nearest or next-nearest etc.) neighbor, as can be realized, e.g., by Rydberg blockade \cite{lukin2001dipole,schachenmayer2010dynamical,pohl2010dynamical,lesanovsky2011many,lauer2012transport,mattioli2015classical,bernien2017probing}.

The dynamics of the QXP contact process on a semi-infinite chain is governed by the Hamiltonian
\begin{align}
    \label{eq:qxp_hamiltonian}
    H_\mathrm{qxp} =
    \sum_{j=1}
    \lambda_{j} \,
    \Bigl(
    \hat Q_{j-1} \hat X_{j} \hat P_{j+1}
    +
    \hat P_{j-1} \hat X_{j} \hat Q_{j+1}\Bigr) + \delta_j \hat Q_j, 
\end{align}
where $\hat Q_j = \ket{\bullet}_{j  j}\!\bra{\bullet}$, $\hat P_j = \ket{\circ}_{j j}\!\bra{\circ}$, $\hat X_j = \ket{\circ}_{j  j}\!\bra{\bullet} + \ket{\bullet}_{j  j}\!\bra{\circ}$, and we allow for a site-dependent coherent excitation rate $\lambda_j$ and an on-site energy $\delta_j$.
It has been shown 
that adding spontaneous decay at rate $\gamma$ from the excited to the ground state, and assuming homogeneous excitation rates, also this (quantum) model shows an absorbing-state phase transition at some critical value of $\lambda/\gamma$ depending on the spatial dimension
\cite{marcuzzi2016absorbing,carollo2019critical}. Although in dimensions $d=2$ and higher the transition is believed to be of second order and of DP universality, the character of the transition in $d=1$ remains not fully understood, but is likely not of the DP universality class \cite{marcuzzi2016absorbing,buchhold2017nonequilibrium,gillman2019numerical,carollo2019critical}.

We here first numerically simulate the coherent contact process starting from an initial seed excitation at site $j=1$
for the case of alternating excitation rates $\lambda_{j=v,w}$ (see Fig.~\ref{fig:intro}a)). A classical spreading process, whose dynamics is fully incoherent and is described by rate equations, as detailed in the Supplemental material,
would simply lead to a diffusive spreading
as shown in see Fig.~\ref{fig:intro}c). 
The quantum case is dramatically different. E.g., if the excitation rate at the odd sites $\lambda_v$ is much smaller than that at the even sites $\lambda_w$, and the lattice is finite with an even number of sites, one observes an oscillation of the excitation between only one site - the initial seed site - and the entire lattice. This is evident in Fig.~\ref{fig:intro}d).
As we will show, this dynamic occurs because the network on which the spreading process takes place corresponds to the Su-Schriefer-Heeger (SSH) model \cite{ssh_model}, which 
possesses topologically protected edge states. 
By periodically modulating on-site energies and excitation rates in time, we will show that one is able to control the spreading of excitations by means of a topological Thouless pump \cite{thouless1983}.

The QXP model, studied in our work, can be realized through coherent Rydberg facilitation in a chain of neutral atoms trapped in optical tweezers. A laser field couples the atoms between their ground $\ket{g}$ and a high-lying Rydberg state $\ket{r}$. 
In the Rydberg state, the atoms interact via a strong van-der-Waals potential, i.e., ${V(r) \sim r^{-6}}$, which falls off very rapidly with distance. Therefore, it can be ignored beyond nearest neighbor distances.
The resulting many-body Hamiltonian, realizing model \eqref{eq:qxp_hamiltonian}, is given by
\begin{align}
    \label{eq:ryd_hamiltonian}
     H_\mathrm{ryd} = 
    \sum_{j=1}^N \Bigl(
    \lambda_j \hat \sigma_j^x
    + \Delta_j \hat n_j
    + V_\mathrm{NN} \hat n_j \hat n_{j+1}\Bigr),
\end{align}
with $\hat \sigma_j^x = \ket{g}_{j j}\!\bra{r} + \ket{r}_{j j} \!\bra{g}$, the projector onto the Rydberg state of the $j$th atom ${\hat n_j = \ket{r}_{j j} \!\bra{r}}$, laser detuning $\Delta_j$, Rabi frequency $\lambda_j \propto \lambda_0$, and interaction potential $V_\mathrm{NN}$ between neighboring Rydberg atoms.

By choosing laser parameters such that $|\Delta_j| \gg \lambda_0$ for all $j$, the driving becomes far off-resonant and individual excitations are suppressed. Furthermore, the lattice spacing $a_0$ is chosen to fulfil the \textit{facilitation constraint}, ${V(a_0) + \Delta = 
V_\mathrm{NN}+\Delta=0}$. Under this condition, to good approximation, atoms are only resonantly driven in the presence of exactly \textit{one} neighboring Rydberg atom, as required by the QXP model.

To simulate the dynamics of Eq.~\eqref{eq:ryd_hamiltonian}, we employ time-evolving block decimation (TEBD) \cite{vidal2003efficient,daley2004time}. For the presented results we use parameters ${\Delta / \lambda_0 = -500}$, with ${V_\mathrm{NN} + \Delta = 0}$ throughout this work. 
We note that for resonant driving, i.e. $\Delta=0$, the interaction between Rydberg atoms results in a full suppression of excitations surrounding a single Rydberg atom, i.e. Rydberg blockade. This is the situation required for the realization of the PXP model \cite{schachenmayer2010dynamical,pohl2010dynamical,lesanovsky2011many,lauer2012transport}, which is hence a special case of the more general model we study here.
%

\paragraph{Domain model of quantum contact process --}
%
Starting from Hamiltonian~\eqref{eq:qxp_hamiltonian}, we realize 
that the many-body dynamics are kinetically constrained, such that excitations can only be created in the presence of one excited neighbor (cf.~Fig.~\ref{fig:intro}). 
We consider a finite one dimensional lattice with the initial (seed) state $\ket{\psi_{m=1}} \equiv \ket{\bullet \circ \circ...\circ}$ and an even number of sites $N$.
Consequently, the many-body dynamics reduces to the dynamics of domains of contiguous excited sites. The domains can grow and shrink at their edges, but they can neither coalesce nor split. 
Remarkably, for the case of a single initial seed excitation, corresponding to a domain size $m=1$,
the mapping of Eq.~\eqref{eq:ryd_hamiltonian} onto the dynamics of the domain size allows to describe the physics in terms of a simple effective single-particle model \cite{marcuzzi2017}: 
\begin{align}
    H_\mathrm{dom} = 
    \sum_{m=2}^{N}
    \lambda_{m} 
    (
    \hat 
    c_{m}^\dagger 
    \hat
    c_{m-1}
    +\mathrm{h.c.}
    ) + \sum_{m=1}^N \eta_m\hat c_m^\dagger \hat c_m.
    \label{eq:dom_hamiltonian}
\end{align}
Here, the creation and annihilation operators $\hat c_m^{(\dagger)}$ represent raising and lowering operators of the domain size $m$. Within Eq. \eqref{eq:dom_hamiltonian}, they take the role of hard core bosons, which can tunnel with hopping rates $\lambda_m$ between sites. These rates relate to the site-dependent excitation rate as $\lambda_{j=m}$ (cf.~Fig.~\ref{fig:intro}), and $\eta_m=\sum_{j=1}^m \delta_j$.
This mapping to the domain space is unique if and only if the first atom is not driven, i.e., $\lambda_1=\delta_1=0$. 

%
\begin{figure}[H]
  \begin{center}
  \includegraphics[width=\columnwidth]{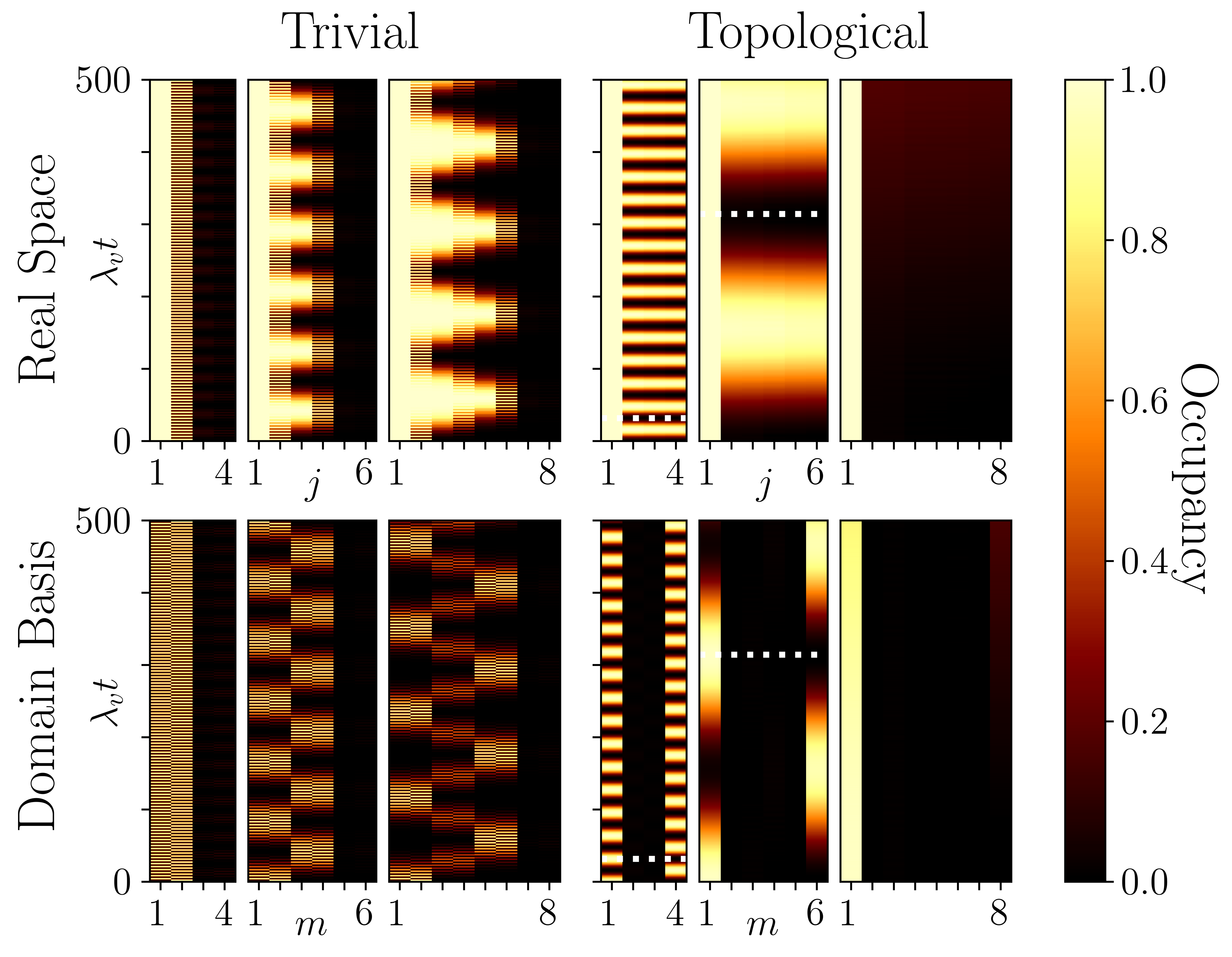}
  \end{center}
  \caption{\textit{Time resolved dynamics}. Top row: real space, bottom row: domain space,
  for the trivial $\lambda_w = 1/10< \lambda_v = 1$ (left column) and topological cases $\lambda_w = 10 > \lambda_v = 1$ (right column). The dotted lines correspond to the analytically expected period $T_\mathrm{hyb} \propto (E_+ - E_-)^{-1}$, where $E_\pm$ are the corresponding energies of the two perfectly localized edge states.}
\label{fig:ssh_dynamics}
\end{figure}
%
\paragraph{Su-Schrieffer-Heeger lattice --}
Considering a lattice with an even number of sites and allowing for alternating flip-rates $\lambda_m$, with
\begin{align}
    \label{eq:lambda}
    \lambda_m = 
    \begin{cases}
        \lambda_v = \lambda_0 = 1, \; \mathrm{even} \; m
        \\
        \lambda_w, \; \mathrm{odd} \; m   
    \end{cases},
\end{align}
where $\lambda_0 = 1$ is setting the energy-scale. 
We can thus create the Su-Schrieffer-Heeger (SSH) model \cite{ssh_model}, a simple model system featuring topological properties. 

The SSH model with open boundary conditions is topologically non-trivial if
$\lambda_w > \lambda_v$. For even number of lattice sites $N$ and $N \to \infty$
it manifests two exponentially localized edge states with localization length ${\xi = ({\log | \lambda_w | - \log | \lambda_v |})^{-1}}$. 

In Fig.~\ref{fig:ssh_dynamics}, we show the time resolved dynamics of the excitation spreading of the QXP process in the SSH lattice starting from the single excited site $j=1$.
The simulations are performed for the many-body Rydberg system Eq.~\eqref{eq:ryd_hamiltonian}, and subsequently mapped onto the SSH model.

Both, the topologically trivial case $\lambda_w < \lambda_v$ (left column) as well as the non-trivial one (right column) 
are considered  in the strong localization limit $\xi < 1$. 

The dynamics exhibit a striking difference for the two cases. In the topological trivial case, where no edge states exits, the excitation simply spreads coherently with rates $\lambda_{v,w}$ until it encounters the boundary, followed by a coherent de-facilitation cascade propagating to the left. When the initial site $j=1$ is reached, the process repeats. This can be understood in  the domain picture, where it corresponds to the oscillation of a wavefunction of a single particle initially prepared in 
a superposition of bulk states. 
In the topologically non-trivial case, the dynamics are completely different. There is an oscillation  between a state where only site $j=1$ is excited and a state where all sites are excited, corresponding to edge-state oscillations in the SSH model. The period of this oscillation rapidly increases with system size $N$ as the two edge states hybridize for finite $N$ with a small energy splitting 
$ E_\pm = \pm |n_1 \mathrm{e}^{-\frac{(N-1)}{\xi}} \lambda_v|$,
where $n_1$ is the population of the edge site \cite{asboth2016short}.
For $\lambda_w \gg \lambda_v$ an initial excitation at $j=1$ approximately prepares the system in an edge state. The second edge mode corresponds to all sites being excited $\ket{\bullet ... \bullet}$, hence $m = N$.
As a result of the hybridization,
the system then oscillates between the  edge states  with period ${T_\mathrm{hyb} = \frac{2 \pi}{E_+ - E_-}}$, shown as dotted white lines in Fig.~\ref{fig:ssh_dynamics}, assuming $n_1=1$.

\begin{figure}[H]
  \centering
  \includegraphics[width=\columnwidth]{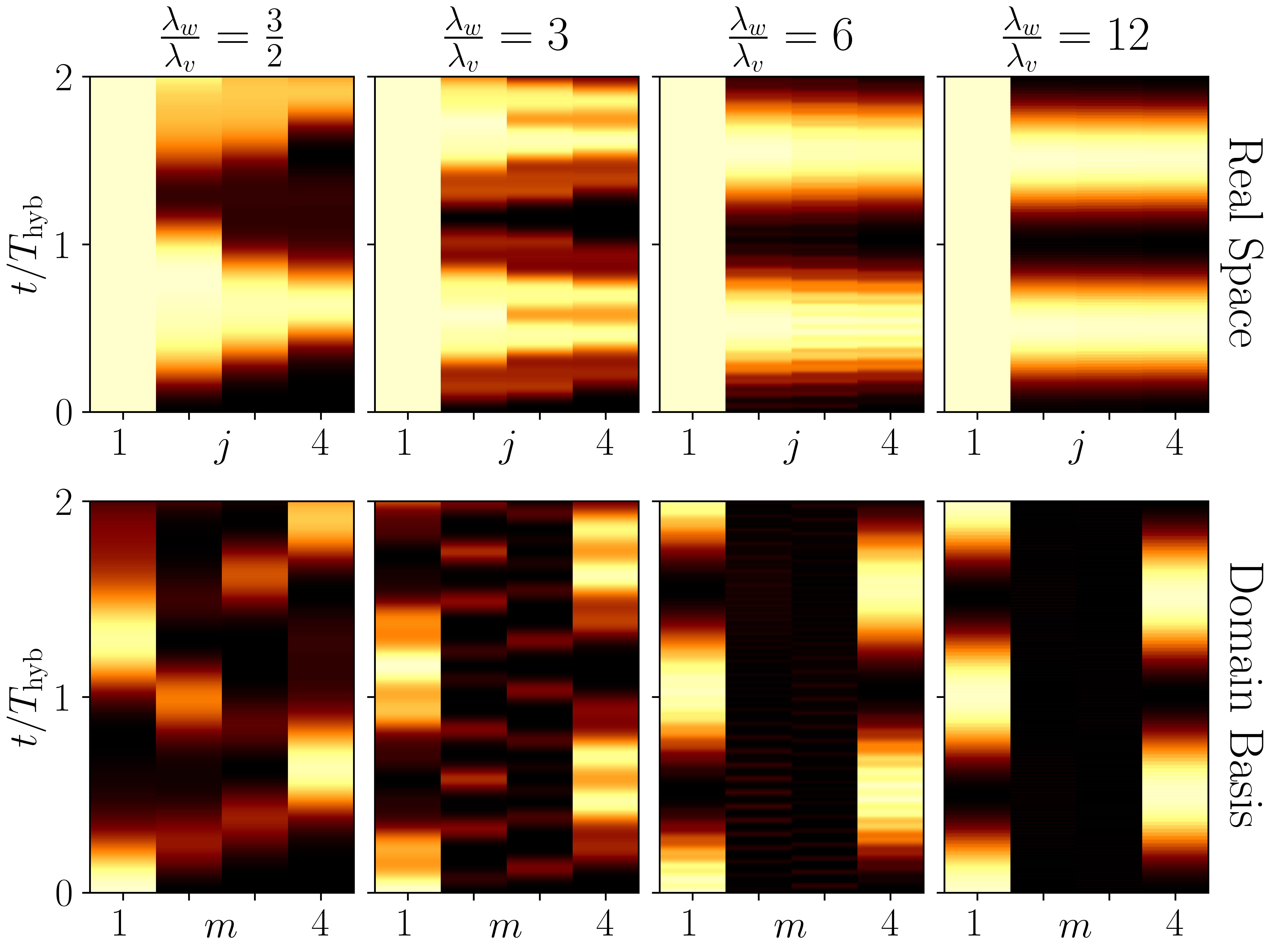}
  \caption{\textit{Hybridization of edge states.} The states $\Psi_L = \ket{\bullet \circ \circ ... \circ}$ and $\Psi_R = \ket{\bullet \bullet \bullet ... \bullet}$ correspond to topological edge states in the limit $\lambda_w / \lambda_v \to \infty$. As the ratio $\lambda_w / \lambda_v$ increases, oscillations between $\Psi_L$ and $\Psi_R$ become clearer and the frequency approaches the analytical expectation $T_\mathrm{hyp}$ (see main text). Color code as in Fig.~2.}
\label{fig:fidelity}
\end{figure}

The domain states $\vert \psi_{m=1}\rangle$ (only one excited site) and $\vert \psi_{m=N}\rangle$ (the whole lattice is excited) correspond to the edge states only for very different hopping rates $\lambda_w$ and $\lambda_v$. As shown in Fig.~\ref{fig:fidelity} many more domain states are 
involved for smaller ratios, and only if $\lambda_w / \lambda_v$ is increased, the edge state oscillations in the domain basis become much clearer and the oscillation period converges to the expected value for perfect localization.

%
%

\paragraph{Coherent control of excitation spreading --}
\label{sec:coherent_control}

The mapping onto the SSH model reveals that the system can be manipulated and controlled using techniques borrowed from topological systems. To demonstrate this, we now turn to enacting topological control over the system dynamics via a Thouless pump \cite{thouless1983}. This allows us to precisely control the domain size and with that the spread of excitations.
To this end, we add a time-dependent on-site potential $\eta_m(t)$ in the domain basis and allow for time-dependent hopping rates $\lambda_m(t)$, 
\begin{align}
    H_\mathrm{dom} = 
    \sum_{m=2}^{N}
    &\lambda_{m}(t)
    (
    \hat 
    c_{m}^\dagger 
    \hat
    c_{m-1}
    +\mathrm{h.c.}
    )
    + \eta_m(t)
    \hat 
    c_{m}^\dagger 
    \hat
    c_{m}.
    \label{eq:effective_AAH_hamiltonian}
\end{align}
The values of $\lambda_m(t)$ and $\eta_m(t)$ are chosen periodic in $m$ with lattice period $q$ such that $H_\mathrm{dom}$ represents 
the topological (time-dependent) Aubry-Andr\'e-Harper (AAH) model \cite{aubry1980analyticity,harper1955single} with $q$ bands (for $N\to\infty$). 
Time-periodic modulation of the parameters leads to a quantized transport of the center of mass of a single-particle wavefunction, governed by a topological invariant (Thouless pump) \cite{citro2023thouless}.
Due to the finite widths of the single-particle bands, the spatial wavefunction will however disperse. This spread, which for the contact process would result in blurring the local excitation probability, can be suppressed by following the high-order tunneling suppression protocol from Ref.~\cite{lee2019}. Specifically, we here consider a model with a unit-cell size $q=3$ and the parameters:
\begin{subequations}
    \begin{align}
        \lambda_m(t) &= \lambda_0  \sin \left( \omega  t + \frac{4\pi}{3} (m + 1) \right),\\
        \eta_m(t) &= \eta_0  \cos \left(\omega  t + \frac{4\pi}{3} (m + 1) \right),
    \end{align}
    \label{eq:effective_parameters_aah}
\end{subequations}
with $\lambda_0 = 1$ again setting the energy-scale. 
Most importantly, the dispersion suppression requires a strong on-site potential, which we realize by choosing $\eta_0 = -10  \lambda_0$. 

In the Rydberg system we correspondingly choose a detuning $\Delta_j(t) = \Delta_0 + \delta_j(t) = -V_\mathrm{NN}+\delta_j(t)$, with
\begin{align}
    \nonumber
    \delta_j(t) &= \eta_{j}(t) - \eta_{j-1}(t) = -\sqrt{3} \eta_0 \sin \left(\omega  t + \frac{4\pi}{3}j + \frac{2\pi}{3}\right)
\end{align}
for $j \geq 2$ (see Supplementary 
for details).

In Fig.~\ref{fig:control} we show the real-space time-evolution of the QXP-chain,
obtained by TEBD simulations of the Rydberg system. The initial seed starting at site-index $j = 1$ facilitates its neighbor controlled by the underlying Thouless-dynamics. To check whether the facilitated domain is not a superposition of different sizes, the lower plot shows a projection of the many-body-state onto the domain-space, with the projection operator $\hat P_m = \ket{m}\bra{m}$, demonstrating perfect matching with Fock states even after several pump cycles. 
Here, a pumping process is shown, in which the domain is first grown with constant rate. It is then kept at the same size, subsequently reduced again, before finally kept at the same size for a short period by 
modifying the frequency $\omega$ through the system parameters as displayed in Fig. \ref{fig:control}a). 
We emphasize that the topological control of the spreading dynamics in the QXP model requires that the many-body dynamics, Eq.\eqref{eq:ryd_hamiltonian}, can be faithfully mapped to the single-particle domain model. If this is 
%
\begin{figure}[H]
  \centering
  \includegraphics[width=\columnwidth]{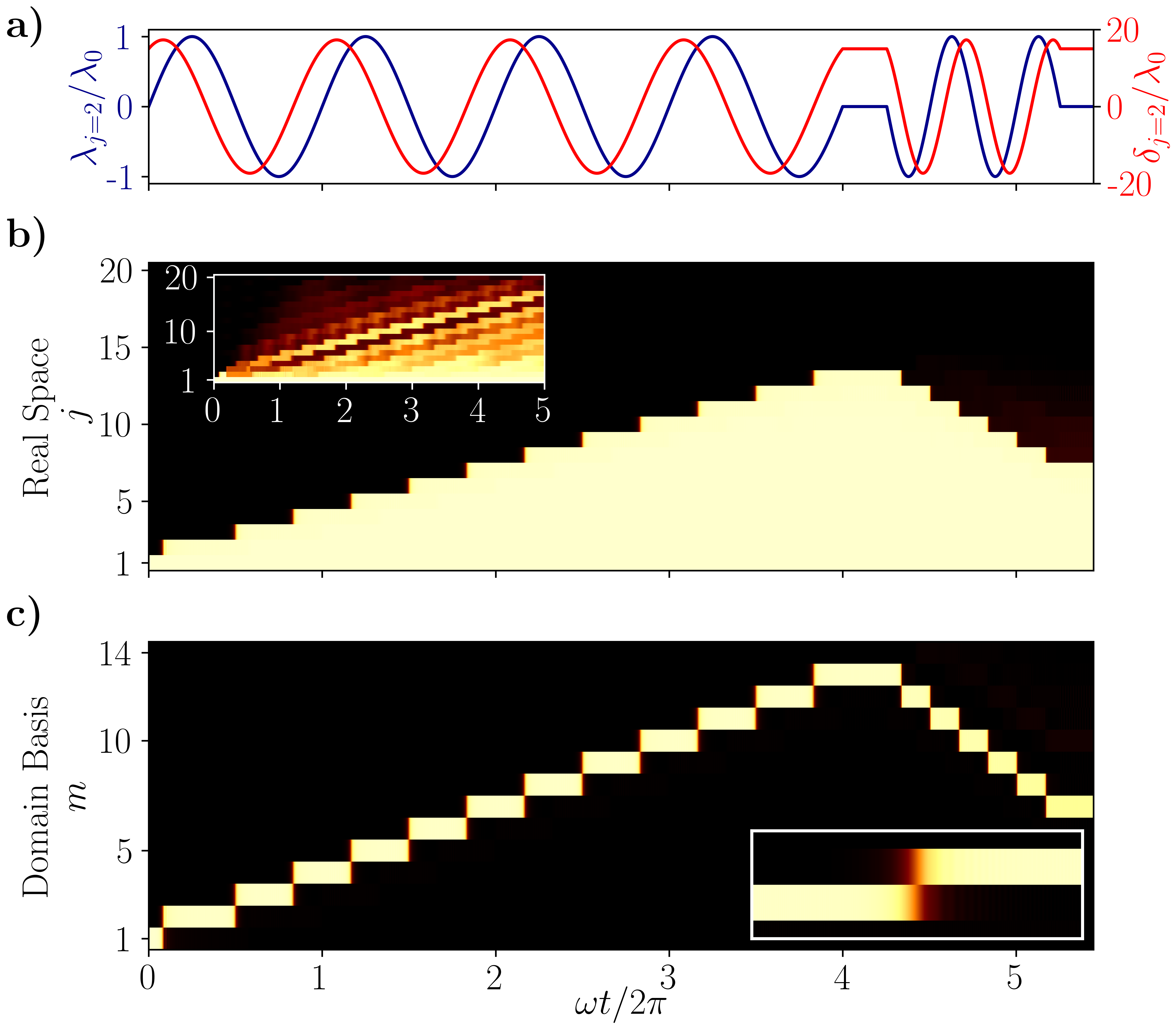}
  \caption{\textit{Quantum control of domain size by manipulating AAH topological pump.} In (a) the corresponding parameters $\lambda_2(t)$ and $\delta_2(t)$ are shown. (b) and (c) show the dynamics in real space and domain basis, respectively.
  The inset in (b) shows 3 periods of a time-evolution, with small detuning $\Delta_0 / \lambda_0 = -22$, where the mapping to the single-particle domain model fails.
  The inset in (c) shows the rapid transition between the domain sizes. 
  To ensure the adiabaticity of the time-evolution, the frequency is chosen as $\omega = 0.02 \lambda_0$.
  }
  \label{fig:control}
\end{figure}
%
\noindent not the case (see inset of Fig. \ref{fig:control}b)), the control of the spreading process is lost. 

Our results show that employing coherent manipulation and topological properties of the network on which coherent spreading occurs, the domain size of excitations can be completely controlled by the underlying Thouless physics.
There are, however, only two limiting factors:
(i) Because of non-adiabatic effects, when the lattice is switched to become static, a minimum time in the static scenario is necessary before turning on again the pump in either direction.
(ii) The slope of the pumping process is limited due to adiabaticity requirements of the underlying Thouless pump.
The latter is determined by the band gap of the effective single-particle AAH model, which is on the order of $\lambda_0$, and thus we require for the frequency of the Thouless pump $\omega \ll \lambda_0$.

\paragraph{Summary and conclusion --}

We have shown that the dynamics of quantum contact processes on regular networks 
can be much richer than its classical counterpart. In particular, if the network is topologically non-trivial, the spreading can be controlled and this control is robust due to topological protection. We here have demonstrated such a topological control for a particular quantum contact process, the QXP model (which is an idealization of coherent Rydberg facilitation) for a single initial seed excitation on a one-dimensional lattice and the quantum counterpart to classical epidemic models. The constrained many-body dynamics of this model can be mapped to an effective single-particle lattice model where the size of the contiguous domain of excited Rydberg atoms corresponds to the position of
a particle in the lattice.
For a finite 1D network with alternating coherent excitation rates, which, in the domain picture, corresponds to the topological Su-Shrieffer-Heeger (SSH) lattice, we have shown that coherent excitation spreading can be confined to a topologically protected subspace.  Here oscillations occur between states where either one edge site or all sites are excited, if the SSH lattice is in the
non-trivial phase and the excitation rates are sufficiently different. Since edge states are topologically protected by generalized symmetries, this confinement of the spreading dynamics is robust against small perturbations which do not close the energy gap of the SSH model or destroy its symmetries. Extending the SSH model to an Andre-Aubry-Harper model with time-periodic hopping amplitudes, implements a Thouless pump with controlled and quantized growth of the excitation domain in the quantum contact process. Our work demonstrated that in contrast to classical processes, the 
dynamics of quantum contact processes can be significantly affected by  
topological properties of the underlying networks, offering entirely new avenues of control of spreading processes.

\paragraph{Acknowledgments --}

Financial support from the DFG through SFB TR 185, Project No. 277625399, is gratefully acknowledged by J.B., M.F. and D.B.. R.S. acknowledges support by the DFG under Germany's Excellence Strategy EXC 2181/1 - 390900948 (the Heidelberg STRUCTURES Excellence Cluster), and CRC1225 ISOQUANT, project-ID 273811115.
The authors also thank the Allianz f\"ur Hochleistungsrechnen (AHRP) for giving us access to the “Elwetritsch” HPC Cluster.

\bibliographystyle{apsrev4-2}
\bibliography{references}

 \newpage
 \onecolumngrid
\section{SUPPLEMENTARY}
\label{sec:supplementary}

\subsection{Rate equation approach for classical contact process}

The classical contact process in a one-dimensional lattice, corresponding to the constrained QXP dynamics, is described by rate equations for the excitation probability $p_j$ of the $j$th 
site:
\begin{equation}
    \frac{d}{dt} p_j = -\bigl(\Gamma_f +\gamma\bigr) p_j + \Gamma_f\, (1-p_j)
\end{equation}
where $\gamma$ is the rate of spontaneous decay and $\Gamma_f$ the facilitation rate that is constrained by the excitation probability of the neighboring sites
\begin{equation}
    \Gamma_f =  \Gamma_f^0\cdot \Bigl[p_{j-1} (1-p_{j+1}) + (1-p_{j-1}) p_{j+1}\Bigr].
\end{equation}
For the Rydberg facilitation considered here it holds $\Gamma_f^0= 2 \Omega^2/\gamma_\perp$, where $\gamma_\perp$ is the dephasing rate of the ground-Rydberg transition.

\subsection{Comparison of QXP and Rydberg Dynamics}

In the main text, the many-body dynamics of coherent Rydberg facilitation was mapped to an effective single-particle Hamiltonian.
 Here we derive the parametrization of the topological pump in the Rydberg chain. In this context, we give some details on the validity limits of this mapping.

The many-body Hamiltonian for the Rydberg facilitation reads
\begin{align}
    \label{eq:ryd_hamiltonian_supplemental}
     H_\mathrm{ryd} = 
    \sum_{j=1}^N \Bigl(
    \lambda_j \hat \sigma_j^x
    + \Delta_j \hat n_j
    + V_\mathrm{NN} \hat n_j \hat n_{j+1}\Bigr).
\end{align}
with $\hat \sigma_j^x = \ket{g}_{j  j}\! \bra{r} + \ket{r}_{j j}\! \bra{g}$, the projector onto the Rydberg state of the $j$th atom ${\hat n_j = \ket{r}_{j j}\! \bra{r}}$, the site-dependent laser detuning $\Delta_j=\Delta_0+\delta_j$, and coherent excitation rate (Rabi frequency) $\lambda_j$, and the van-der-Waals interaction potential $V_\mathrm{NN}$ between neighboring Rydberg atoms. The time dependency is suppressed for notational simplicity.
We assume a large offset detuning, i.e.,
\begin{equation}
    \vert \lambda_j\vert, \vert \delta_j\vert \ll \vert \Delta_0\vert, \label{eq:fac-condition-1}
\end{equation}
and consider nearest-neighbor facilitation conditions, i.e. choose $\Delta_0$ such that 
\begin{equation}
    V_\textrm{NN} + \Delta_0 =0. \label{eq:fac-condition-2}
\end{equation}
Conditions \eqref{eq:fac-condition-1} and \eqref{eq:fac-condition-2}
imply that the excitation of an atom into the Rydberg state is suppressed unless there is exactly one nearest neighbor already in the Rydberg state. This defines the QXP model, eq.\eqref{eq:qxp_hamiltonian}, of the main text.

For a (quantum) contact process there has to be an initial seed.
Here we assume that initially there is only a single excited atom that sits on a boundary site and is not driven, i.e., $\lambda_1 = 0$ and $\Delta_1 = 0$. Then one can describe the system with an effective domain Hamiltonian
\begin{align}
    \nonumber
    H_\mathrm{dom} = 
    \sum_{m=2}^{N}
    \lambda_{m}
    (
    \hat 
    c_{m}^\dagger 
    \hat
    c_{m-1}
    +\mathrm{h.c.}
    )
    + \eta_m
    \hat 
    c_{m}^\dagger 
    \hat
    c_{m},
\end{align}
with $m$ characterizing the length of the domain of atoms excited to the Rydberg state. Note, that if the initial excitation would be in the bulk, the mapping to the domain size would be ambiguous.
The on-site potential $\eta_m(t)$ in the single-particle cluster model originates from the on-site potential of the Rydberg atoms and their interaction potential as follows:
\begin{align}
    \eta_m = \sum_{j=2}^m \Bigl(\Delta_j + V_\mathrm{NN}\Bigr)= \sum_{j=2}^N \delta_j.\label{eq:eta_directly_to_rydberg}
\end{align}

For the topological pump, the effective hopping amplitudes and on-site potentials are chosen periodic in space and time according to
\begin{subequations}
    \begin{align}
        \lambda_m(t) &= \lambda_0  \sin \left( \omega  t + \frac{4\pi}{3} (m + 1) \right),\\
        \eta_m(t) &= \eta_0  \cos \left(\omega  t + \frac{4\pi}{3} (m + 1) \right),
    \end{align}
    \label{eq:supplemental_parametrization_single_particle}
\end{subequations}
which ensures the suppression of the dispersion of the effective particle at all times if $\eta_0 = 10 \lambda_0$ is chosen.



\begin{figure}[H]
    \centering
    \includegraphics[width=\columnwidth]{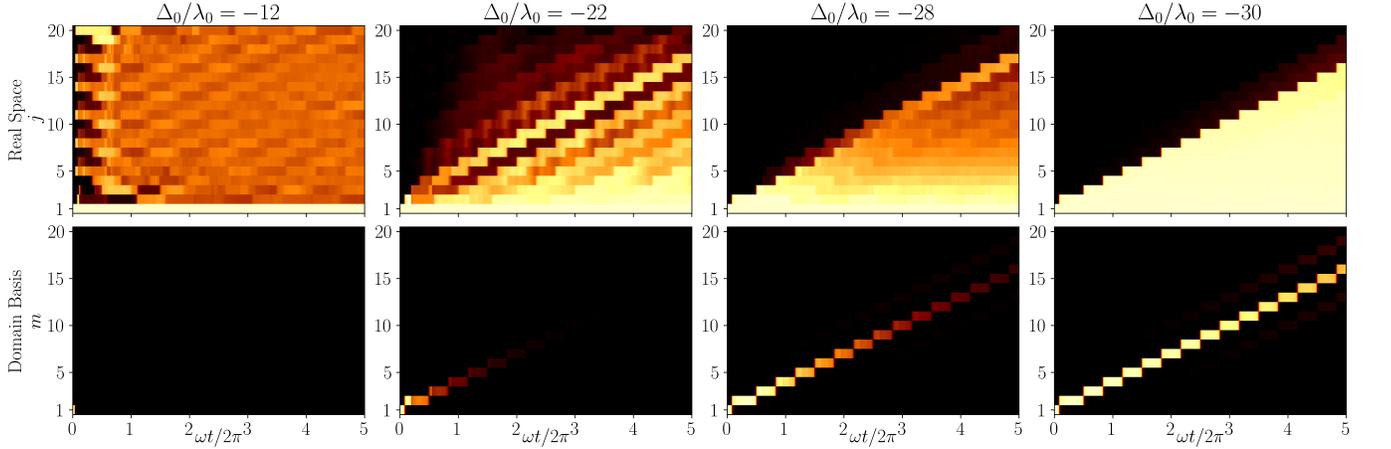}
    \caption{\textit{Accuracy of the single-particle QXP model.} For smaller detunings $\Delta_0/\lambda_0$ the QXP model of the facilitation breaks down. \textit{Top}: Real space Rydberg excitation probability. \textit{Bottom}: Population in domain picture. Color code is the same as in Fig.~2 of the main text.}
    \label{fig:accuray_of_effective_model}
\end{figure}

The following choice of $\delta_j$ satisfies Eq.~\eqref{eq:eta_directly_to_rydberg}:
\begin{align}
    \delta_j &= \eta_{m=j} - \eta_{m=j-1}
    = -\sqrt{3} \eta_0 \sin \left(\omega  t + \frac{4\pi}{3}j +\frac{2\pi}{3}\right).
\end{align}
Given this, the maximum amplitude of the variation is $\delta_\mathrm{max} = \sqrt 3 \eta_0 \approx 17 \lambda_0$. To ensure the facilitation regime for all times, the offset $\Delta_0$ has to be large compared to $\delta_\mathrm{max}$.
To underline this, the results of simulations for different values of $\Delta_0$ are shown in Fig. \ref{fig:accuray_of_effective_model}. 
As can be seen, values below the critical amplitude are dominated by individual excitations. The periodicity of the excitations in space are simply a consequence of the spatial periodicity of the detunings. As soon as $\Delta_0$ becomes of the order of $\delta_\mathrm{max}$, the dynamics changes towards the one expected by the QXP model.

Since individual excitations and the facilitation process do not exclude each other, there is no sudden transition to the facilitation regime, when increasing $\Delta_0$. Individual excitations have an amplitude $\propto \frac{\lambda_0^2}{\lambda_0^2 + \Delta_j^2}$, that vanishes for large detunings $\Delta_j$.

Throughout this manuscript, we ensured the facilitation regime by setting $\Delta_0 = 500 \lambda_0$, exceeding the critical value $\delta_\mathrm{max}$ by more than one order of magnitude.

%

\end{document}